\documentclass{ifacconf}

\usepackage{graphicx}      % include this line if your document contains figures
\usepackage{natbib}        % required for bibliography
\usepackage{xcolor}
\usepackage{listings}
\usepackage{url}
\usepackage{amsmath, amssymb}
\usepackage{bm}
\newcommand{\condbar}{\,|\,}  % The vertical bar in f(y | x)
\newcommand{\reals}{\mathbb{R}}
\newcommand{\ws}{\;}

\usepackage{fancyhdr}
\usepackage{nopageno}

\begin{document}
\begin{frontmatter}

\title{When is gray-box modeling advantageous for virtual flow metering?} 

\author[First]{M. Hotvedt} 
\author[First,Second]{B. Grimstad} 
\author[Third]{D. Ljungquist}
\author[First]{L. Imsland}

\address[First]{Engineering Cybernetics Department, NTNU, Trondheim, Norway (e-mail: \{mathilde.hotvedt, lars.imsland\}@ntnu.no)}
\address[Second]{Solution Seeker (e-mail: bjarne.grimstad@solutionseeker.no)}
\address[Third]{TechnipFMC (e-mail: Dag.Ljunquist@technipfmc.no)}

\begin{abstract}                % Abstract of not more than 250 words.
Integration of physics and machine learning in virtual flow metering applications is known as gray-box modeling. The combination is believed to enhance multiphase flow rate predictions. However, the superiority of gray-box models is yet to be demonstrated in the literature. This article examines scenarios where a gray-box model is expected to outperform physics-based and data-driven models. The experiments are conducted with synthetic data where properties of the underlying data generating process are known and controlled. The results show that a gray-box model yields increased prediction accuracy over a physics-based model in the presence of process-model mismatch. They also show improvements over a data-driven model when the amount of available data is small. On the other hand, gray-box and data-driven models are similarly influenced by noisy measurements. Lastly, the results indicate that a gray-box approach may be advantageous in nonstationary process conditions. Unfortunately, choosing the best model prior to training is challenging, and overhead on model development is unavoidable. 
\end{abstract}


\end{frontmatter}
%===============================================================================

\section{Introduction}\label{sec:introduction}
\thispagestyle{fancy}
\chead{\textit{2021 M. Hotvedt, B. Grimstad, D. Ljungquist, L. Imsland. This work has been submitted to IFAC for possible publication}}
\pagenumbering{gobble}
Gray-box modeling is a methodology that integrates physics-based modeling with machine learning techniques in process model development \citep{Willard2020}. The gray-box models are placed on a grayscale dependent on the degree of integration, ranging from physics-based to data-driven models. A common perception is that physics-based models require little data in development and are more robust to noisy measurement than data-driven models. This perception arguably stems from the high extrapolation capabilities demonstrated by many physics-based models \citep{Oerter2006}. Nevertheless, complex physical phenomena may be challenging to model in detail using first principles, and simplifications are generally necessary to make them suitable for real-time control and optimization applications \citep{Roscher2020}. Simplifications reduce the model capacity and thereby the ability to capture arbitrarily complex physical behavior. Therefore, physics-based models often have a bias, or process-model mismatch \citep{Hastie2009}.

In contrast, many data-driven models have a large capacity, typically reducing model bias. Furthermore, some data-driven models are computationally cheap to evaluate and are therefore suitable for real-time applications. Moreover, they commonly have lower development and maintenance costs compared to physics-based models \citep{Solle2016}. On the other side, due to the inherent bias-variance trade-off \citep{Hastie2009}, a large capacity often results in high variance. High variance causes data-driven models to struggle with extrapolation to future process conditions and to yield low performance in the small data regime \citep{Roscher2020}. Gray-box modeling is expected to leverage the complementary and advantageous properties of physics and data to minimize both bias and variance. In other words, create a model that achieves high performance in the presence of process-model mismatch, little or noisy data, which extrapolates well to previously unseen process conditions and is computationally efficient. From the data-driven domain point of view, gray-box modeling is similar to introducing strong priors in the model. In image classification using convolutions neural networks, strong priors, here in terms of parameter sharing, resulted in state-of-the-art performance \citep{Hastie2009}.

One application where accurate process models are of high importance is virtual flow metering (VFM). A VFM is a soft-sensor able to predict the multiphase flow rate in real-time at convenient locations in a petroleum asset \citep{Toskey2012}. The standard practice in the industry today is physics-based models, and several commercial simulators exist \citep{Amin2015}. In later years, data-driven VFM models have demonstrated high performance \citep{AlQutami2017a, AlQutami2017b, AlQutami2017c, ALQutami2018, Bikmukhametov2019, Grimstad2021}. On the other hand, due to the inherently complex multiphase flow rate characteristics and that the available data typically resides in the small data regime \citep{Grimstad2021}, gray-box VFMs have gained increasing attention, see \citep{Bikmukhametov2020a, Hotvedt2020a, Hotvedt2020b, Hotvedt2021} and references therein. However, superior performance over physics-based or data-driven models has yet to be demonstrated. This article contributes in this direction by investigating four scenarios where a gray-box approach is believed to excel and offer higher performance than non-gray-box alternatives. These are formulated as four hypotheses: 

\subsubsection{Hypothesis 1}  Under mismatch between a physics-based VFM and the process, a gray-box VFM developed from the physics-based VFM achieves higher performance. 

\subsubsection{Hypothesis 2} 
With little available data, a gray-box VFM obtains higher performance than a data-driven VFM.  

\subsubsection{Hypothesis 3} 
Increasing the noise level on the data, a gray-box VFM is less influenced than a data-driven VFM.  

\subsubsection{Hypothesis 4} In nonstationary conditions, a gray-box VFM yields higher performance than a data-driven VFM.  

In Hypothesis 1, the increased capacity of the gray-box compared to the physics-based model is believed to be significant. In Hypothesis 1-3, the decreased capacity of the gray-box compared to the data-driven model is believed to be decisive as it may reduce model variance. In real life, available process data can have several uncontrolled characteristics, for instance, faulty sensor measurements. Such characteristics make it challenging to examine and conclude on the hypotheses as it is unknown whether a poor model performance results from the modeling technique or the available data. Therefore, in this work synthetic data designed to explore the hypotheses are generated by a simulator of a petroleum production choke. In several idealized experiments, the properties of gray-box production choke models are compared to physics-based and data-driven models. The remaining article is structured in the following way. The simulator is introduced in Section \ref{sec:simulator}, descriptions of the synthetic datasets are given in Section \ref{sec:data}, and the VFM models are presented in Section \ref{sec:models}. Thereafter, the experiments are described and results visualized in Section \ref{sec:case-study}, the results are discussed in Section \ref{sec:discussion}, and concluding remarks are given in Section \ref{sec:concluding-remarks}. 

\section{The simulator}\label{sec:simulator}
\begin{figure}[ht]
\centering
\includegraphics[width=0.5\columnwidth]{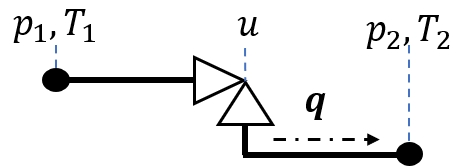}   % printed width is 8.4
\caption{Illustration of the production choke valve and typically available measurements.} 
\label{fig:choke}
\end{figure}

The simulator is a physics-based petroleum production choke valve model. A typical production choke along with available measurements is illustrated in Fig. \ref{fig:choke}.
The multiphase mass flow rate (a mixture of oil, gas, and water) $\bm{\dot{m}} = (\dot{m}_{\text{oil}}, \dot{m}_{\text{gas}}, \dot{m}_{\text{water}})$ through the choke restriction is calculated using an advanced version of the Sachdeva model \citep{Sachdeva1986}, where slip effects, allowing the gas and liquid phases to move with unequal velocity, are included in the model. The slip model is taken from \citep{Alsafran2009}. The model requires measurements of the pressure upstream ($p_1$) and downstream ($p_2$) of the choke valve, the upstream temperature ($T_1$), the choke opening ($u$), and the mass fraction of the phasic fluids $\bm{\eta} = (\eta_{\text{oil}}, \eta_{\text{gas}}, \eta_{\text{water}})$. The mass fractions are assumed to sum to one. Often, the volumetric multiphase flow rate is of interest $q = q_{\text{oil}} + q_{\text{gas}} + q_{\text{water}}$ and can be obtained from the mass flow rates using the mass fractions and fluid densities at standard conditions (SC) \citep{ISO13443}:
\begin{equation}
    q_i = \frac{\eta_i\dot{m}}{\rho_{i,SC}}, \quad i \in \{\text{oil}, \text{gas}, \text{water}\}. 
\end{equation}
In the simulator, an area function relates the choke opening to the effective flow area through the choke $A(u)$. This function will mimic an equal percentage valve, where an equal increment in choke opening results in an equal percentage changed area. The simulator, or process, is referred to as $\mathcal{P}$ and defined by the notation:
\begin{equation}\label{eq:process-cond}
\begin{aligned}
     y &= f(\bm{x}; \bm{\phi}) + \varepsilon \in \reals,
\end{aligned}
\end{equation}
where the model output is the volumetric multiphase flow rate $y=q$, $f$ is the first principle equations, the input measurements are $\bm{x} = (p_1, p_2, T_1, u, \eta_{\text{oil}},\eta_{\text{water}}) \in \reals^6$, and $\phi$ are constant model parameters. Noise is added to $q$ by sampling $\varepsilon$ from a noise distribution, for instance a Gaussian distribution.  

\section{Dataset generation}\label{sec:data}
Process $\mathcal{P}$ in Section \ref{sec:simulator} is used to generate three different datasets $\mathcal{D}_k = \{(\bm{x}_t, y_t)\}_{t=1}^{N_k},\; k=\{1,2,3\}$. The index $t$ reflects time. The three datasets are designed to investigate the four hypotheses from Section \ref{sec:introduction}. The sequence of observations in each dataset is sampled from the joint probability distribution of $\mathcal{P}$: $p_t(\bm{x}, y) = p_t(y \condbar \bm{x}) p_t(\bm{x})$, Here, $p_t(\bm{x})$ is the marginal distribution of the input measurements. Output measurements $y_t$ follow the conditional distribution $p_t(y \condbar \bm{x})$ expressed with \eqref{eq:process-cond}. Notice, $\mathcal{P}$ is allowed to be nonstationary resulting in $p_{t_1}(\bm{x}, y) \neq p_{t_2}(\bm{x}, y)$ for $t_1 \neq t_2$. In this study, nonstationarity is introduced with virtual drift only \citep{Ditzler2015}, meaning that $p_{t_1}(\bm{x}) \neq p_{t_2}(\bm{x})$ for $t_1 \neq t_2$. Real drift resulting in $p_{t_1}(y \condbar \bm{x}) \neq p_{t_2}(y \condbar \bm{x})$ for $t_1 \neq t_2$ is neglected by keeping $f$ unchanged and the parameters $\bm{\phi}$ in $\mathcal{P}$ constant.

Dataset $\mathcal{D}_1$ is generated as a best-case scenario to fairly examine Hypothesis 1-3 in Section \ref{sec:introduction}. Firstly, the process is assumed stationary: $p_{t_1}(\bm{x}) = p_{t_2}(\bm{x}) \forall t$. Secondly, the $\bm{x}$ are independently drawn. This is idealized as measurements in real data are often strongly correlated \citep{Hotvedt2021}. Thirdly, a large range of common process conditions through the lifetime of a petroleum well is covered by sampling the input observations $\bm{x}_t$ from:
\begin{equation}
\begin{aligned}
    p_1 &\sim \mathcal{U}(30, 70) \ws bar, \\
    p_2 &\sim \mathcal{N}(22, 0.5) \ws bar, \\
    T_1 &\sim \mathcal{N}(50, 2) \ws ^{\circ}C\\
    u &\sim \mathcal{U}(0, 100) \ws \%, \\
    \eta_{\text{oil}} &\sim \mathcal{U}(0, 80) \ws\%,\\
    \eta_{\text{water}} &\sim \mathcal{U}(0, 20) \ws\%. \\
\end{aligned}    
\end{equation}
for any $t$. To ensure a sufficient dataset size $N_1 = 10 000$ observations are sampled. Lastly, only normally distributed noise $\varepsilon\sim \mathcal{N}(0, \sigma_{\varepsilon}^2)$ is considered. The included noise levels are $\sigma_{\varepsilon} \in \{1, 2, 3, 4, 5, 10\}$, which results in a coefficient of variation of $\sigma_{\varepsilon}/\mu\in \{0.02, 0.05, 0.07, 0.1, 0.12, 0.24\}$, where $\mu$ is the mean of the noise-free flow rate measurements. Normally distributed noise is an idealized case as measurement sensors may comprise different noise types. The dataset is randomly separated into a training and a test dataset with $N_{1,\text{test}} = 2000$. From the training dataset, $20\%$ are randomly extracted as a validation dataset.

\begin{figure}
\centering
\includegraphics[width=\columnwidth]{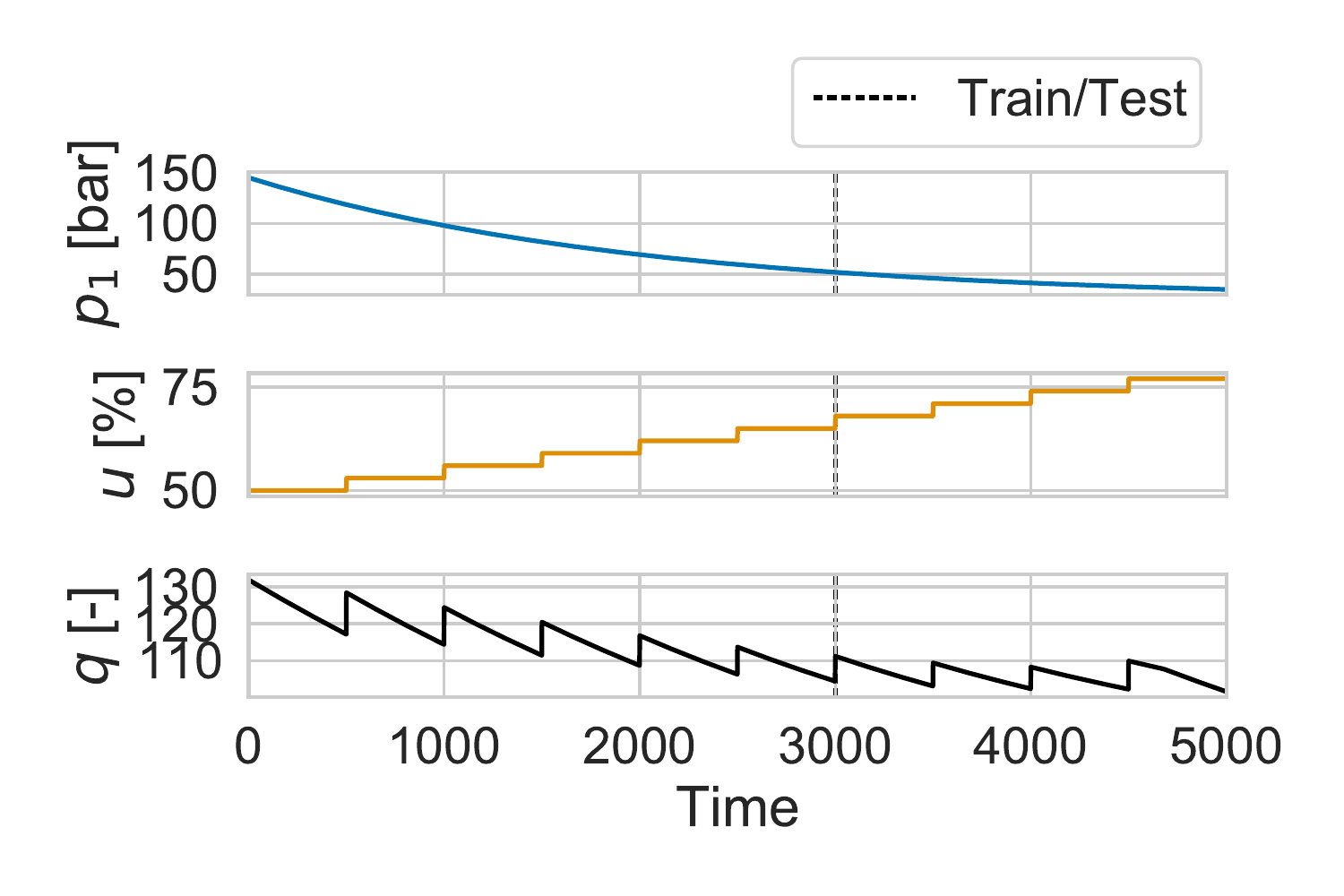}   % printed width is 8.4
\caption{Illustration of the dataset mimicking typical behavior when the reservoir is depleted with time.} 
\label{fig:dataset-depleting-reservoir}
\end{figure}

\begin{figure}
\centering
\includegraphics[width=\columnwidth]{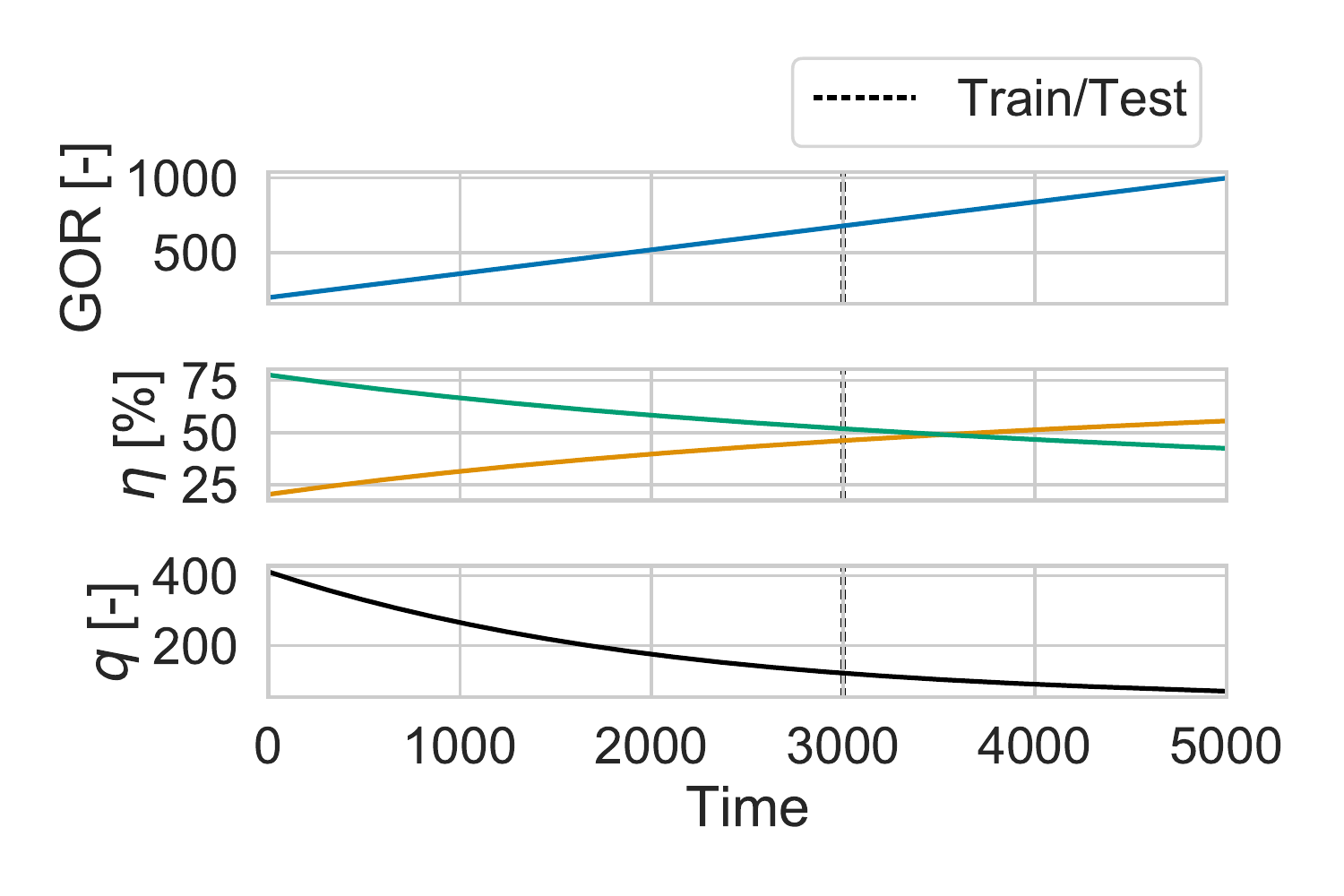}   % printed width is 8.4
\caption{Illustration of the dataset mimicking typical behavior when the gas-to-oil ratio increases. The mass fractions of oil and gas are the green and orange curve, respectively.} 
\label{fig:dataset-GOR}
\end{figure}

The $\mathcal{D}_2$ and $\mathcal{D}_3$ mimics two typical real case scenarios where the process is nonstationary. In both datasets, $N_2 = N_3 = 5000$ noise-free observations are sampled. The datasets are split into training and test according to time with $N_{2,\text{test}} = N_{3,\text{test}} = 2000$. Hence, the models will be used to predict future process responses. The validation dataset is also extracted considering time and consists of the 600 latter training observations. Dataset $\mathcal{D}_2$ mimics a scenario where the reservoir is depleted with time. As a result, the pressure in the reservoir and the upstream part of the choke decreases with time. If the petroleum asset is producing on plateau, process engineers typically increase the choke opening to maintain high production rates \citep{Jahn2008}. This scenario is illustrated in Figure \ref{fig:dataset-depleting-reservoir}. 
The $p_1$ is decreased in time using an exponential function, whereas the choke opening is increased in steps of $2.5$\%. The remaining variables are kept constant for any $t$: $p_2 = 22$ bar, $T_1 = 50^{\circ}C$, $\eta_{\text{oil}} = 85$\%, and $\eta_{\text{water}} = 2$\%. Dataset $\mathcal{D}_3$ mimics a scenario where the gas-to-oil ratio (GOR) increases with time. This phenomenon typically occurs when the reservoir pressure drops below the bubble point pressure such that the gas dissolved in the oil starts to escape \citep{Jahn2008}. Fig. \ref{fig:dataset-GOR} illustrates the resulting flow $q$ and the mass fractions of oil $\eta_{\text{oil}}$ (green) and gas $\eta_{\text{gas}}$ (orange) when the GOR is linearly increased from 200 to 1000. The $p_1$ is the same as for $\mathcal{D}_2$ illustrated in Fig. \ref{fig:dataset-depleting-reservoir}. The remaining variables are kept constant for any $t$: $p_2 = 22$ bar, $T_1 = 50^{\circ}C$, $u = 100$\%, and $\eta_{\text{water}} = 2$\%.

\section{Models}\label{sec:models}
Four production choke models with different capacities have been developed: one physics-based, one data-driven, and two gray-box models. The models will be described briefly below. More details may be found in \citet{Hotvedt2021}. 

The physics-based model is the Sachdeva model, referred to as M, and defined by the short notation 
\begin{equation}
\begin{aligned}
    \hat{y}_{\text{M}} &= f_{\text{M}}(\bm{x}; \bm{\phi}_{\text{M}}) \in \reals,
\end{aligned}
\end{equation}
The true area function is kept unknown, and a linear relationship is utilized instead. Among the $\bm{\phi}_{\text{M}}$ is the discharge coefficient, which is a multiplicative calibration factor used to change the magnitude of the area function. In industrial VFMs, additional calibration factors exist to change the shape of the function. Here, these are excluded to restrict the capacity of M, enforcing a significant mismatch between $\mathcal{P}$ and M.  

The data-driven model is a fully connected, feed-forward neural network. Naturally, other types of data-driven models could be used instead, yet, the neural network is selected due to its large capacity. The model D is defined by 
\begin{equation}
\begin{aligned}
    \hat{y}_{\text{D}} &= f_{\text{D}}(\bm{x}; \bm{\phi}_{\text{D}}) \in \reals,
\end{aligned}
\end{equation}
where $\bm{\phi}_{\text{D}} =  \{(\bm{W}_1, b_1),\ldots (\bm{W}_L, b_L)\}$ are the weights and biases in the neural network on each layer $l=1,...,L$. The rectified linear unit is used as activation function.

The gray-box models are based on the M. Two variants are examined. The first is an error model where a data-driven model attempts to capture additive mismatches between $\mathcal{P}$ and M. This model is referred to as H-E:
\begin{equation}
\begin{aligned}
    \hat{y}_{\text{H-E}} &= f_{\text{H-E}}(\bm{x}; \bm{\phi}_{\text{H-E}}) \\
    & = f_{\text{M}}(\bm{x}; \bm{\phi}_{\text{M}}) + f_{\text{D}}(\bm{x}; \bm{\phi}_{\text{D}}) \in \reals.
\end{aligned}
\end{equation}
The second hybrid model addresses the unknown area function of $\mathcal{P}$ by multiplying the initial linear function of the M with a neural network: $A = A_{\text{M}}\times A_{\text{D}}$. The effect will be that the magnitude and shape of the area function may be adjusted. This model is referred to as H-A:
\begin{equation}
\begin{aligned}
    \hat{y}_{\text{H-A}} &= f_{\text{H-A}}(\bm{x}; \bm{\phi}_{\text{H-A}}) = f_{\text{M}}(\bm{x}, A_{\text{D}}; \bm{\phi}_{\text{M}})\in \reals\\
    A_{\text{D}} & =f_{\text{D}}(\bm{x}; \bm{\phi}_{\text{D}}) \in \reals.
\end{aligned}
\end{equation}
As the neural network in H-A is multiplied with a small value ($A_{\text{M}}$), the capacity of the H-A is likely smaller than the capacity of H-E. This can be argued by acknowledging that large outputs from the network in H-A will be less influential on the flow rate predictions than a large output from the network in H-E. Additionally, the advanced Sachdeva model used for $\mathcal{P}$ has been implemented to examine the process-model mismatch. This model will be referred to as M$^*$ and differ from $\mathcal{P}$ by keeping the true $\bm{\phi}$ unknown. Hence, any process-model mismatch will be a consequence of parameter deviation away from the true value.

For all models the $\bm{\phi}$ are estimated using maximum a posteriori (MAP) estimation:
\begin{equation}\label{eq:map}
\begin{aligned}
    \bm{\phi}^{\star}_{MAP} &= \arg \max_{\bm{\phi}} p(\bm{\phi} \condbar \mathcal{D}_k) \\
    &= \arg \min_{\bm{\phi}} \Big[\sum_{i=1}^{N_k} \frac{1}{\sigma_{\varepsilon}^2}\left(y_i - \hat{y}\right)^2 \\
    &+ \sum_{i=1}^{m} \frac{1}{\sigma_{i}^2}\left(\phi_i - \mu_i\right)^2\Big]. 
\end{aligned}
\end{equation}
where $m$ is the number of parameters. The priors on $\bm{\phi}$ are assumed normal $\phi \sim \mathcal{N}(\mu, \sigma^2 )$. The optimization problem is solved using stochastic, iterative, gradient-based optimization with the optimizer Adam \citep{adam} and early stopping, a common approach in the data-driven domain. Details of the training approach are given in \citet{Hotvedt2021}. 

\section{Case study}\label{sec:case-study}
Four experiments (Exp. 1-4) have been conducted to answer the four hypotheses in Section \ref{sec:introduction}. Below, each experiment will be described, and the results visualized. Due to stochasticity, the experiments are run several times, called trials. The results of the trials will be visualized in figures with the median ($p_{50}$) as a solid line and a shaded area to indicate the lower ($p_{25}$) and upper ($p_{75}$) quantiles. 

\subsection{Exp. 1 - decreasing dataset size}
\subsubsection{Description} This experiment examines the performance of the models to a decreasing training dataset size. Dataset $\mathcal{D}_1$ is used for this purpose using the noise-free measurement of the flow rate. The considered training data lengths are $N \in \{2, 4, 8, 20, 40, 80, 800, 4000, 8000\}$. The training data is randomly extracted from $\mathcal{D}_1$ in each trial. 

\subsubsection{Results}
\begin{figure}
\centering
\includegraphics[width=\columnwidth]{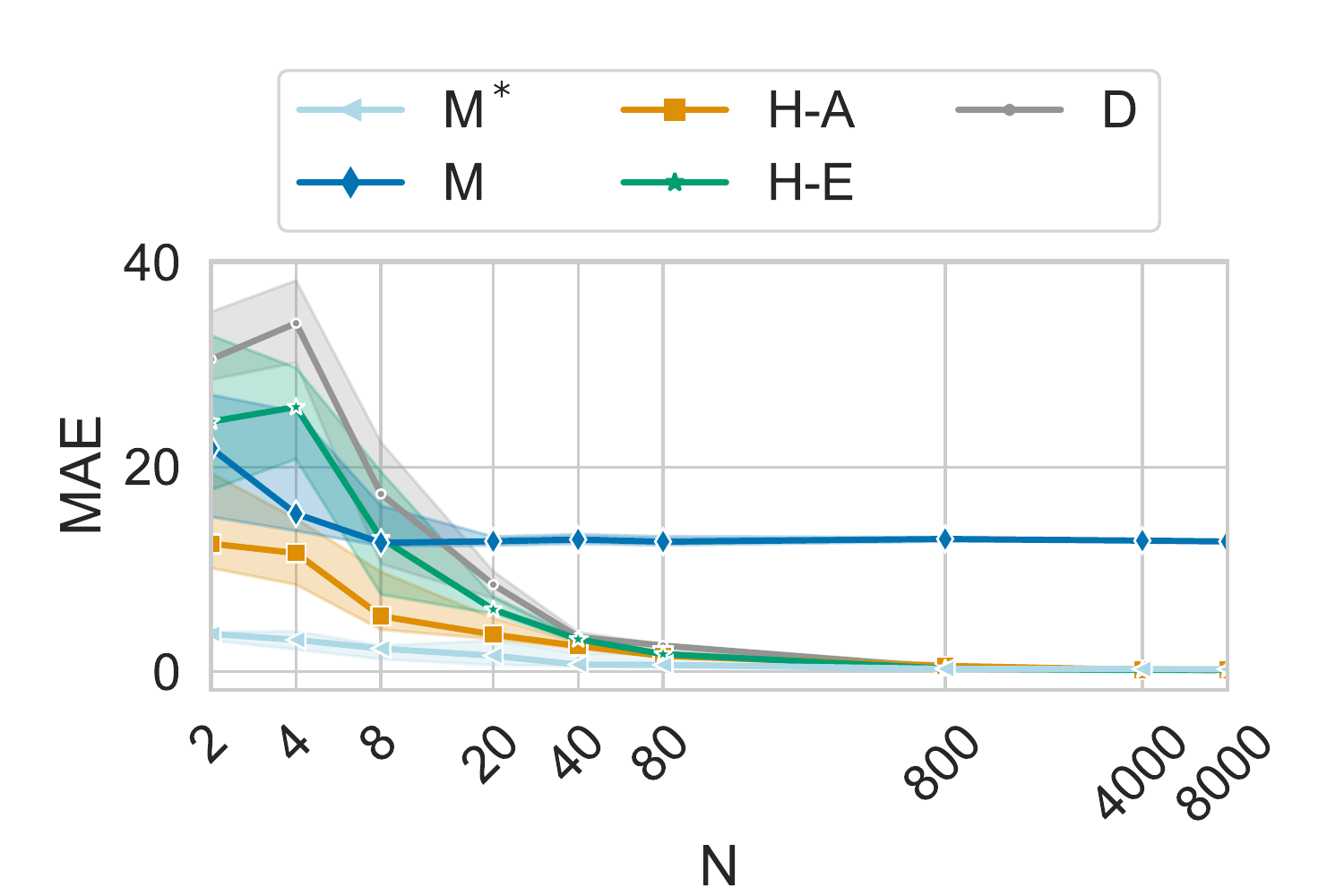}   % printed width is 8.4
\caption{The mean absolute error as a function of the training set size.} 
\label{fig:error-vs-train-size}
\end{figure}

The model performance in terms of the mean absolute error (MAE) is visualized as a function of N in Fig. \ref{fig:error-vs-train-size}.

\subsection{Exp. 2 - increasing noise level}
\subsubsection{Description} This experiment investigates the robustness of the models to an increased noise level. The models will be trained using dataset $\mathcal{D}_1$ and the output measurements with the different noise levels $\sigma_{\varepsilon}$ in Section \ref{sec:data}, one at a time. The performance at different noise levels is calculated using the noise-free measurements as the basis of comparison. In other words, treating the noise-free $q$ as the true value. 

\subsubsection{Results}
\begin{figure}
\centering
\includegraphics[width=\columnwidth]{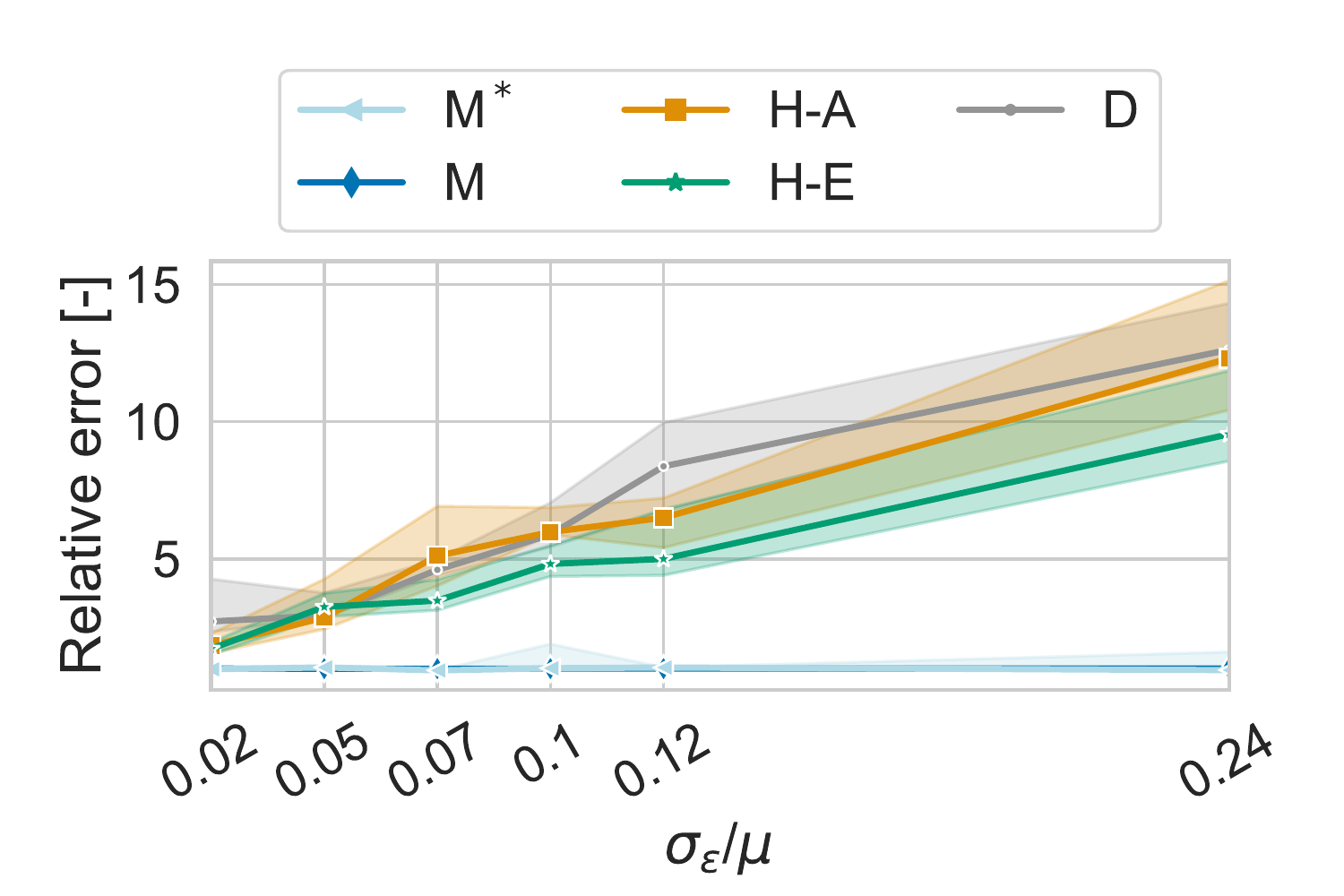}   % printed width is 8.4
\caption{The relative error as a function of the coefficient of variation for the models.} 
\label{fig:error-vs-noise-level}
\end{figure}

Fig. \ref{fig:error-vs-noise-level} shows the relative error of the models as a function of the coefficient of variation $\sigma_{\varepsilon}/\mu$. The relative error is calculated by dividing the MAE obtained at one noise level by the MAE obtained with noise-free measurements. A relative error larger than $1.0$ means the model performance has decreased.

\subsection{Exp. 3 - the depleting reservoir}
\subsubsection{Description} Dataset $\mathcal{D}_2$ is used to analyze the model performances in the nonstationary case of a depleting reservoir. 
\subsubsection{Results}
\begin{figure}
\centering
\includegraphics[width=\columnwidth]{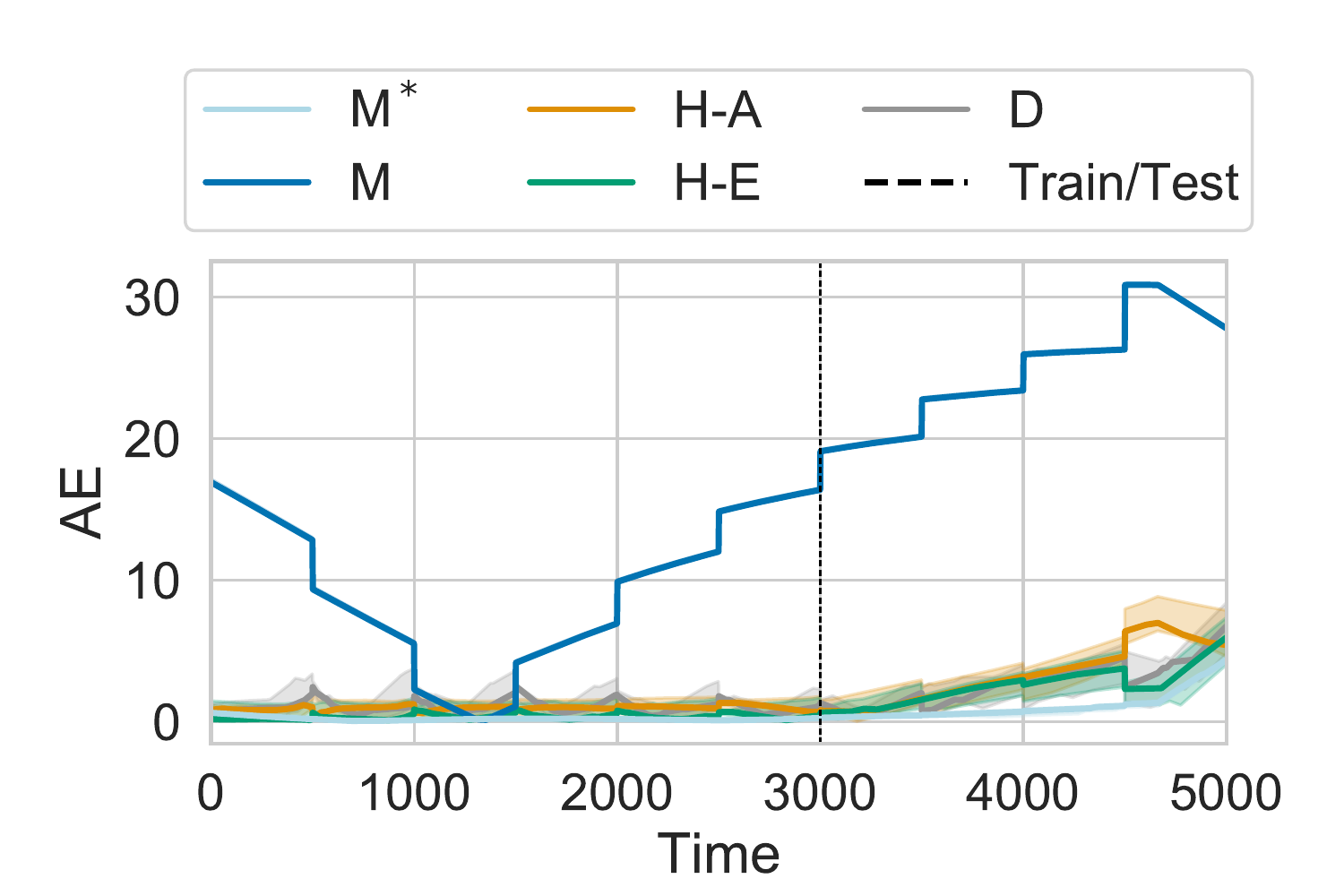}   % printed width is 8.4
\caption{The absolute error of the model predictions as a function of time for Exp. 3.} 
\label{fig:time-series-p1-u-ae}
\end{figure}

The absolute value of the prediction error (AE) in time is visualized for the different models in Fig. \ref{fig:time-series-p1-u-ae}. The black, dotted line separates training and test data. Table \ref{tb:depleting-reservoir} gives the validation and test MAE for the models. 
\begin{table}[ht]
\centering
\caption{The validation and test mean absolute error in Exp. 3.}
\begin{tabular}{rlllll}
& M$^{\star}$ & M & H-A & H-E & D \\ \hline      
MAE$_v$   & 0.1 & 18.8 & 2.2 & 1.3 & 2.5\\
MAE$_t$   & 1.0 & 24.7 & 4.3 & 2.5 & 2.8 \\\hline
\end{tabular}
\label{tb:depleting-reservoir}
\end{table}

\subsection{Exp. 4 - increasing gas-to-oil ratio}
\subsubsection{Description} Dataset $\mathcal{D}_3$ is used to analyze the model performance in the nonstationary case of an increasing GOR.
\subsubsection{Results}
\begin{figure}[ht]
\centering
\includegraphics[width=\columnwidth]{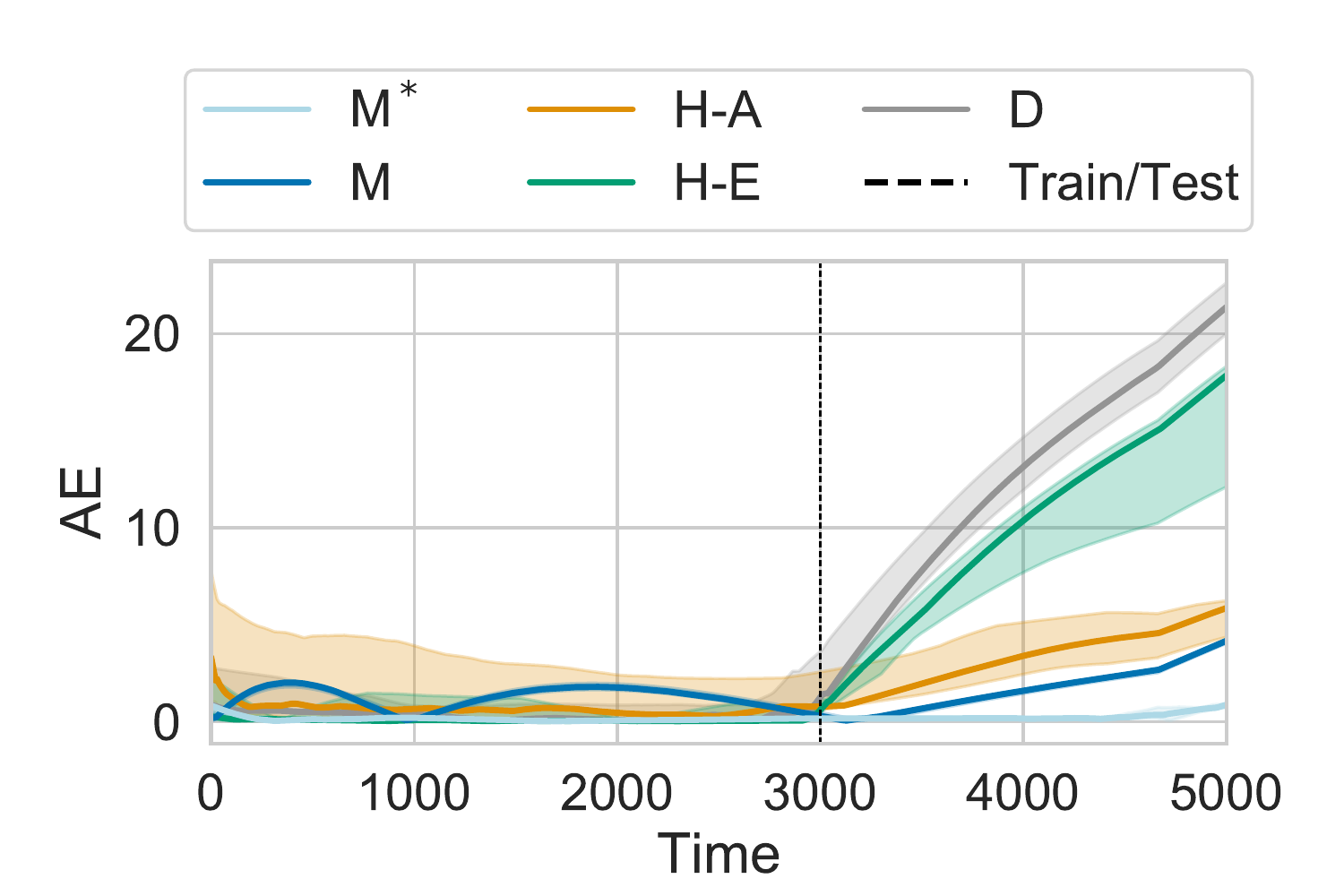}   % printed width is 8.4
\caption{The absolute error of the model predictions as a function of time in Exp. 4.} 
\label{fig:time-series-gor-ae}
\end{figure}

Fig. \ref{fig:time-series-gor-ae} shows the absolute error in time separated into training and test data. Table \ref{tb:increasing-gor} gives the validation and test MAE.  
\begin{table}[h]
\centering
\caption{The validation and test mean absolute error in Exp. 4.}
\begin{tabular}{rlllll}
& M$^{\star}$ & M & H-A & H-E & D \\ \hline      
MAE$_v$   & 0.2 & 2.0 & 3.0 & 4.9 & 6.9 \\
MAE$_t$   & 0.3 & 1.6 & 3.4 & 9.0 & 12.7 \\ \hline
\end{tabular}
\label{tb:increasing-gor}
\end{table}

\section{Discussion}\label{sec:discussion}
Firstly, notice from Fig. \ref{fig:error-vs-train-size} that only M gives a large process-model mismatch for large dataset sizes. This indicates that the other models have a sufficient capacity to remove the bias if exposed to an adequate dataset size. Few observations ($N > 80$) were required for the D and Hs to obtain negligible MAE, which suggests that the process is simple to learn. With increasingly complex processes, a larger number of observations would likely be required to remove the bias. Secondly, Fig. \ref{fig:error-vs-train-size} show that the error increases the most for the D when the dataset size decreases. This imply that the D has the largest variance and adapts much to the training data, thus decreasing the generalizability to the unobserved test data. Further, Fig. \ref{fig:error-vs-train-size} indicates that the H-E has a larger variance than the H-A due to a larger increase in MAE for decreasing N. This is likely due to the model architecture as commented in Section \ref{sec:models}. Fig. \ref{fig:error-vs-noise-level} show that the M and M$^{\star}$ are robust against an increasing noise level, whereas the Hs and D are not. This confirms that the Hs and D have a larger variance. On the other hand, Fig. \ref{fig:error-vs-noise-level} shows that the Hs barely achieve a better performance than the D. Moreover, it seems that the H-E has a lower variance than the H-A, which is conflicting with the results in Fig. \ref{fig:error-vs-train-size}. However, H-E is designed to capture additive mismatches, which is the only considered noise influence and may explain the slightly better performance. 

The results from Exp. 1-2 indicates that gray-box models may yield lower variance than a data-driven model and reduce bias in physics-based models. Therefore, in nonstationary conditions the expectation is that the Hs will perform better than the D and better than the M if there are large process-model mismatches. Figs. \ref{fig:time-series-p1-u-ae}-\ref{fig:time-series-gor-ae} and Tables \ref{tb:depleting-reservoir}-\ref{tb:increasing-gor}does show that at least one H performs better than the D in both experiments and that it is advantageous with an H when the process-model mismatch is large as in Exp. 3. The large mismatch in Exp. 3 is a consequence of the available measurements of $u$ making the assumed linear shape of the area function in M of greater influence than in Exp. 4 where $u=100\% \ws \forall t$. It should be noted, the U-shaped curve of the M on the training data in Fig. \ref{fig:time-series-p1-u-ae} is due to the objective function in \eqref{eq:map}, and the performance on the test data can probably be improved by weighing the recent observations the most. 

On the other hand, in Exp. 3, the performance of the D is comparable with the Hs. In Exp. 4, the discrepancy between the Hs is large, where the H-A and H-E yields a good and poor performance, respectively. Consequently, it is challenging to determine which model will perform the best before training. Ideally, the best model could be deduced a priori by examining known process-model mismatches and the capacity of the models. Nevertheless, this showed nontrivial even for these idealized experiments. For instance, in Exp. 3, the H-A was expected to perform best as it targets the discrepancy between the linear and true area function. Nevertheless, H-E yields the best performance, closely followed by the D. Therefore, model selection must be performed posterior to training using the validation dataset. Accordingly, extracting the validation dataset representatively is important, for instance, according to time for nonstationary processes. Positively, the results in Tables \ref{tb:depleting-reservoir}-\ref{tb:increasing-gor} indicate that the errors on the validation data are illustrative for the model performances on the test data as the best model yields the lowest error on both. A disadvantage is that this approach increases the overhead on model development and testing. The observant reader notices that the model performances in Figs. \ref{fig:time-series-p1-u-ae}-\ref{fig:time-series-gor-ae} decrease with time. This is a typical scenario for steady-state modeling in nonstationary conditions. Utilization of learning methods for frequent model updating would likely improve the long-term performances.

\section{Concluding remarks}\label{sec:concluding-remarks}
Overall, the results in this research show that a gray-box approach to VFM may reduce both model bias and variance compared to a physics-based and data-driven approach, respectively. From the results in Section \ref{sec:case-study} and the discussions in Section \ref{sec:discussion}, Hypotheses 1 and 2 are confirmed: a gray-box model performs better than a physics-based model in the presence of process-model mismatch and performs better than a data-driven model when the training dataset size decreases. On the other side, the gray-box and data-driven models have comparable performances for an increasing data noise level and Hypothesis 3 cannot be confirmed. 

The results from two experiments with data from nonstationary process conditions showed that a gray-box model can improve the performance of a data-driven model, hence, confirming Hypothesis 4. Moreover, the gray-box model can significantly improve the performance of a physics-based model in nonstationary conditions if there are large process-model mismatches. On the other hand, the results also show that it is challenging to determine prior to model training, for instance, based on known process-model mismatches, which model yields the best performance in different scenarios. Therefore, the best model must be chosen posterior to training using the error on a validation dataset. Consequently, overhead on model development and testing is unavoidable.

Certainly, the hypotheses were only investigated on synthetic data and generalization to real life is challenging. In real life, there may be other undesired and unknown characteristics of the process complicating model development. For instance, increasingly complex and rare physical phenomena, or heteroscedastic measurement noise. Moreover, this work only considers two scenarios of nonstationary process behavior, although possible scenarios are numerous. Additionally, other gray-box model variants may yield different results in different scenarios. 

Nevertheless, the results from this work indicate that gray-box modeling is advantageous for virtual flow metering if there are large process-model mismatches, little available data, and nonstationary environments. 

\begin{ack}
This research is a part of BRU21 - NTNU Research and Innovation Program on Digital and Automation Solutions for the Oil and Gas Industry (\url{www.ntnu.edu/bru21}) and supported by Lundin Energy Norway.    
\end{ack}

\bibliography{ifacconf}     

\end{document}